\documentclass{aa}

\usepackage{epsf}
\usepackage{amssymb}
\usepackage{amsmath}
\usepackage{amsxtra}
\usepackage{times}
\usepackage{graphicx}

\def\aa{A\&A}

\begin{document}

   \title{Galaxy and hot gas distributions in the z=0.52 galaxy cluster RBS380
   from CHANDRA and NTT observations.}

   \author{Rodrigo Gil-Merino$^1$ and Sabine Schindler$^2$}

   \institute{$^1$Institute f\"{u}r Physik, Universit\"at Potsdam, 
              Am Neuen Palais 10, D-14469 Potsdam, Germany.\\
             \email{rmerino@astro.physik.uni-potsdam.de} \\
              $^2$Institut f\"{u}r Astrophysik, Universit\"{a}t Innsbruck,
              Technikerstr. 25, A-6020 Innsbruck, Austria.\\
             \email{Sabine.Schindler@uibk.ac.at}
             }
		 
   \offprints{R. Gil-Merino}

\date{Received date; accepted date}
\date{Submitted: May, 2003}

\titlerunning{The z=0.52 galaxy cluster RBS380}
\authorrunning{R. Gil-Merino \& S. Schindler}

\abstract{We present CHANDRA X-ray and NTT optical observations of
the distant $z=0.52$ galaxy cluster RBS380 -- the most distant
cluster of the ROSAT Bright Source (RBS) catalogue.
We find diffuse, non-spherically symmetric X-ray emission with 
a X-ray luminosity of $L_X(0.3-10$~keV$)=1.6~10^{44}$ erg/s,
which is lower than expected from the RBS. The reason is a bright
AGN in the centre of the cluster contributing considerably to the
X-ray flux. This AGN could not be resolved with ROSAT. 
In optical wavelength we identify several galaxies belonging to the cluster.
The galaxy density is at least $2$ times higher than expected for such a
X-ray faint cluster, which is  another confirmation of the weak
correlation between X-ray luminosity and optical richness. The
example of the source confusion in this cluster shows how
important high-resolution X-ray imaging is for cosmological research.
}

\maketitle

\section{Introduction}

The galaxy cluster RBS380 is part of a large optical programme to
search for strong gravitationally lensed arcs in X-ray luminous
clusters selected from the ROSAT Bright Survey (RBS, Schwope et
al. 2000), with a predicted probability for arcs of 45\%. In
addition to the optical images X-ray observations are taken in
order to compare masses determined with different methods and to
use the X-ray morphology for lensing models. The main goal of this
project is to combine X-ray and optical information, together with
possible gravitational lensing information, to constrain
cosmological models.

The cluster presented here -- RBS380 -- is after RBS797 (Schindler
et al. 2001) the second cluster for which we have performed a
combined optical and X-ray analysis. The X-ray source RBS380 was
found in the ROSAT All-Sky Survey (RASS, Voges et al. 1996, 1999) and
classified as a massive cluster of galaxies in the RBS. RBS380 is
the most distant cluster of this catalogue.

We present here CHANDRA ACIS-I and NTT SUSI2 observations of the
X-ray cluster RBS380 at $z=0.52$ and coordinates
$\alpha=03~01~07.6$, $\delta=-47~06~35.0$ (J2000).

We find a lower X-ray luminosity than expected from the RBS. The
reason is source confusion in ROSAT data -- the X-ray emission of
the central AGN had been mixed up with cluster emission.

The high galaxy number density in this cluster is in contrast to
its low X-ray luminosity. This is another confirmation that
optical luminosity  is not well correlated with X-ray luminosity,
see e.g. Donahue et al. (2001) or the clusters Cl0939+4713 and
Cl0050-24 for extreme examples of optical richness and low X-ray
luminosity (Schindler \& Wambsganss 1996, 1997; Schindler et al.
1998).

Throughout this paper we use $\rm{H}_0 = 65$ km/s/Mpc, 
$\rm{\Omega_M}=0.3$ and $\rm{\Omega_{\Lambda}}=0.7$.

\section{Data acquisition and reduction}

\subsection{X-ray data reduction}

The cluster RBS380 was observed on October 17, 2000 by the CHANDRA
X-ray Observatory (CXO). A single exposure of 10.3 ksec was
obtained with the Advanced CCD Imaging Spectrometer (ACIS). During
the observations the $2\times2$ front-illuminated array ACIS-I was
active, together with the S0 chip of the ACIS-S $1\times6$ array,
although this last one was not used for the data reduction, since
the expected cluster centre was placed on the ACIS-I array. Each
CCD in the ACIS-I is a $1024\times1024$ pixel array, each pixel
subtending $0''.492\times0''.492$ on the sky, covering a total
area of $16'.9\times16'.9$.

The data were ground reprocessed on February 28, 2001 by the CHANDRA X-ray
Center (CXC). The analysis of these reprocessed data was performed by the
CIAO-2.2 suite toolkit.

As upgraded gainmaps from preprocessing were available, we used
the \emph{acis\_process\_events} tool to improve the quality of
the level\_2 events file. We also corrected for aspects offsets
and removed bad pixels in the field. For that we used the provided
bad pixel file acisf02201\_000N001\_bpix1.fits by the CXC. We
built the lightcurve for the observation period and we searched
for short high backgrounds intervals. We found none, so no data
filtering was needed.

Since we are interested in the diffuse emission of the galaxy
cluster, special attention has to be paid to the removal of point
sources. This extra care is not needed when the count rate is high
enough, since the cluster emission can be seen even without any
processing. If the number of counts from diffuse cluster emission
is low, any not removed point source can induce wrong estimates.
In a broadband ($0.3-10$ keV) image, we applied two different
procedures for the detection of sources: \emph{celldetect} and
\emph{wavdetect}.  The latter uses wavelets of differents scales
and correlates them with the image; the former uses sliding square
cells with the size of the instrument PSF. In general,
\emph{celldetect} works well with well-separated point sources,
although a low threshold selection will obviously overestimate the
number of point sources. On the other hand, \emph{wavdetect} tends
to include some diffuse emission regions as point sources. For
these reasons, a scientific judgment must be applied in order to
decide which regions must be identified as point sources. Using a
sigma threshold of $10^{-6}$ in the \emph{wavdetect} routine, we
found 31 point sources, expecting a probability of wrong
detections of 0.1 in the image. Using
analogous criteria for the \emph{celldetect} routine we found no
significant differences.

The correction for telescope vignetting and variations in the spatial
efficiency of the CCDs was done by means of an exposure map, using the standard
procedure of the CIAO-2.2 package. The exposure map was generated for an
integrated energy distribution peak. The value of the peak was slightly
different depending on the included region. Selecting the whole effective area
of the ACIS-I array, the peak value was 0.7 keV. If the selected area was only
the region covering the central part of the cluster (a circle of radio $1'.5$),
the value of the peak was
then 0.5 keV. We used these two values for the reduction and we could not see
any significant change in the final result.

The background correction was done using a blank field background
set acisi\_C\_i0123\_bg\_evt\_230301.fits provided by the CXC. We
used a blank field instead of a region from the science image,
since one cannot be sure a priori whether a certain region in the
field is free of galaxy cluster emission. The smoothing process
for the final image was done with the \emph{csmooth} CIAO tool and compared 
to the result using the IRAF\footnote{IRAF is
distributed by the National Optical Astronomy Observatories, which
are operated by the Association of Universities for Research in
Astronomy, Inc., under cooperative agreement with the National
Science Foundation.} (Image Reduction and Analysis Facility) 
task \emph{gauss} (using a $\sigma=20$ 
pixels Gaussian) to be sure that no artificial features were created in 
the convolution process. We found no significant differences.
\subsection{Optical data reduction}

The galaxy cluster RBS380 was observed in optical wavelength with
the New Technology Telescope (NTT) in service mode during 
summer 2001. The Superb Seeing Imager-2 (SUSI2) camera was used in
bands V and R. The SUSI2 detector is a 2 CCDs array,
$1024\times2048$ pixels each, subtending a total area on the sky
of $5'.5\times5'.5$ (the pixel size in the $2\times2$ binned mode
is $0.16''/$pixel). In order to be able to avoid the gap between
the two chips during the data reduction process, dithering was
applied.

The data reduction was perfomed with the IRAF
package. A total number of 6 images in R band and 3 in V band in 
very good seeing conditions ($\leq1''$) were used in the analysis.
The exposure time was 760 sec for each image. For each band, after
bias subtraction, a standard flatfielding was not enough to
produce good results, because the twilight flats provided by the
NTT team contained some stars and the scientific images showed
stronger gradients than the flats. A hyperflat (see e.g. Hainaut
et al. 1998) was built to flat-correct the images. We briefly
describe the hyperflat technique here.

To produce a hyperflat we processed separately the provided
twilight flats and the scientific images, although the procedure
will be analogous in both sets. The technique is to smooth
strongly all the bias subtracted and normalized frames (with e.g a
Gaussian $\sigma=100$ pixels). 
The result is then subtracted from the
original frames, so one obtains a very flat background, but still
with stars in the images. Smoothing again the result with a
smaller Gaussian (e.g. $\sigma=20$ pixels) will show all the stars.
One can then mark all these stars in the original frames, median
average them and reject the marked values. Applying this procedure
to the twilight flats set and to the scientific images set, one
obtains a final twilight flat and a final night-sky flat,
respectively. A linear combination of these two yields the final
hyperflat.

Once the images are flatfielded, they can be co-added, resulting
in a deep image of the field and free of chip gaps. Note that the
whole procedure has to be done for each filter.

\begin{figure}[hbtp]
 \centering
 \includegraphics[bbllx=113,bblly=209,bburx=498,bbury=586,width=8.0cm,
                  angle=0,clip=true]{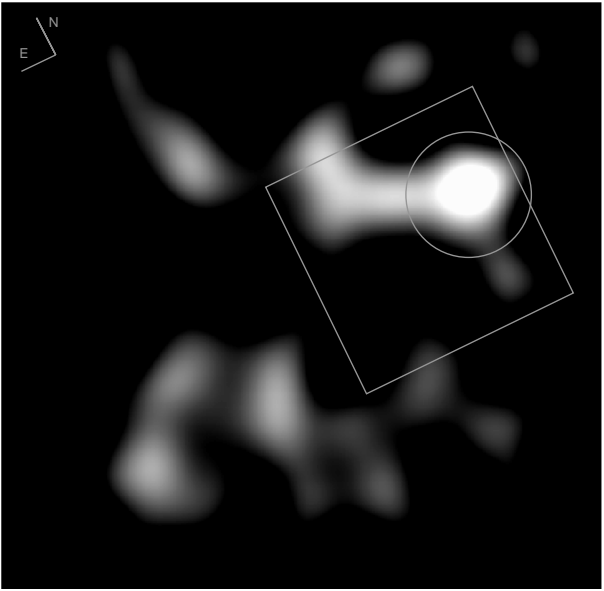}
 \caption[]{X-ray image of RBS380 (z=0.52) in the [0.3-10 keV] band, 
 adaptatively smoothed with the \emph{csmooth} CIAO tool and cross-check 
 with the IRAF \emph{gauss} task. The total area is $14'\times14'$. 
 The rotated square shows the region that was observed in the
 optical band (V and R). The circle with a radius of $1'.5$ marks the
 area within which we have computed a count rate of 0.05 counts/s. Point-like 
 X-ray sources have been removed. North and East are marked.}
 \label{Xcumulo}
\end{figure}
\begin{figure}[hbtp]
 \centering
 \includegraphics[bbllx=118,bblly=77,bburx=475,bbury=418,width=6.0cm,
                  angle=0,clip=true]{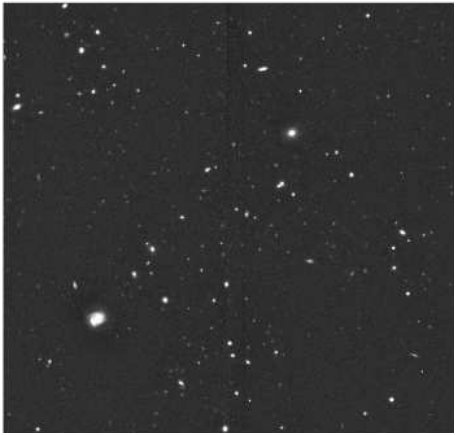}
 \caption[]{Optical R band image of RBS380 (z=0.52). The total area is
 $5'\times5'$. North is up and East is left.}
 \label{Rcumulo}
\end{figure}
\section{Analysis and results}

\subsection{X-ray results}
\label{s_xresults}

The final X-ray image after data reduction (including point sources 
removal) is shown in
Fig.~\ref{Xcumulo}. We encircle the main cluster emission within a
radius of $1'.5$ centred on the peak of the emission. The count
rate obtained in that area is 0.05 counts/s. We compared this count 
rate to the count rate of the same region in the background fields, 
finding a value of 0.02 counts/s. We found that this background count 
rate was in fact not very sensitive to its position in the field, as 
expected. Using a weighted 
average column density $nH=2.23~10^{20}$~cm$^{-2}$ (Dickey \&
Lockman 1990), a Raymond-Smith source model with $T=5$~keV and the
cluster redshift $z=0.52$, the derived luminosity is
$L_X (0.3-10 {\rm{keV}})=1.6~10^{44}$ erg/s. Using slightly lower numbers for the
temperature in the source model (in the range 3-4 keV), reduces
the final luminosity result in only by a few per cent. This is a
relatively low X-ray luminosity for a massive cluster of galaxies.
As the luminosity is so low we were particularly careful with the
background subtraction and the removal of point sources.

The X-ray luminosity is lower than expected from the RBS 
results. The reason is an X-ray point source centred on the
coordinates $\alpha=03~01~07.8$ and $\delta=-47~06~24.0$. The
point source is probably an AGN which could not be resolved with
ROSAT and therefore not distinguished from cluster emission. The
AGN is probably the central cluster galaxy. Within a radius of $7''$
we find a count rate of 0.07 counts/s for this point source. 
Using a power law model
with photon index 2, the same column density as for the cluster
and an energy range [0.3-10~keV], the obtained flux for this AGN
is $f_X=8.2~10^{-13}$ erg~cm$^2$~s$^{-1}$. This AGN is one of the
galaxies for which the RBS optical follow-up observations (Schwope
et al. 2000) yielded a redshift of 0.52 (see Fig.~\ref{Rgal}). 
In Table~\ref{agntabla} we summarise the coordinates, count 
rates and luminosities of the AGN and the cluster.

\begin{table}[tb]
 \centering
 \begin{tabular}{ccccc}
  \hline\noalign{\smallskip}
   Name & $\alpha_{2000}$ & $\delta_{2000}$ & Counts [cts/s] & L$_X$ [erg/s]\\
  \noalign{\smallskip}\hline\noalign{\smallskip}
   AGN    &  03~01~07.8 & -47~06~24.0 & 0.07 & 1.8~10$^{44}$\\
   RBS380 &  03~01~07.6 & -47~06~35.0 & 0.05 & 1.6~10$^{44}$\\
  \noalign{\smallskip}\hline
 \end{tabular}
 \caption{Coordinates of the AGN and the cluster. The AGN is almost at 
 the centre of the cluster emission. We also show the count rate for the 
 two objects (normalized for the different apertures, see text for 
 details) and the luminosities, both in the [0.3-10~keV] band (bolometric 
 luminosity for the cluster is given in Table~\ref{cuadro}). The contribution 
 of the AGN is larger than the cluster luminosity. Both objects are at the
 same redshift of 0.52. An optical counterpart of the AGN is marked in Fig.
 ~\ref{Rgal}.}
\label{agntabla}
\end{table}
In addition to the main cluster emission within a circle of radius
$1'.5$ described above, we found an asymmetric structure extending
to both sides of this main region. If this is cluster emission, it
could indicate that the cluster is not relaxed, but interacting
with surrounding material or/and an infalling galaxy group. In any
case we consider the inferred X-ray luminosity $L_X$ as an lower
limit for the cluster. Due to the low number of X-ray counts we did not
perform any spectral analysis.

\subsection{Optical results}

Both V and R images show a high number density of galaxies. The
main goal is to find a way of selecting the cluster members in
order to determine their number and their spatial distribution. We
select cluster members through a colour-magnitude relation,
applying it to all the galaxies detected both in V and R bands.

\begin{figure}[hbtp]
 \centering
 \epsfxsize=5.2 cm \rotatebox{-90}{\epsffile{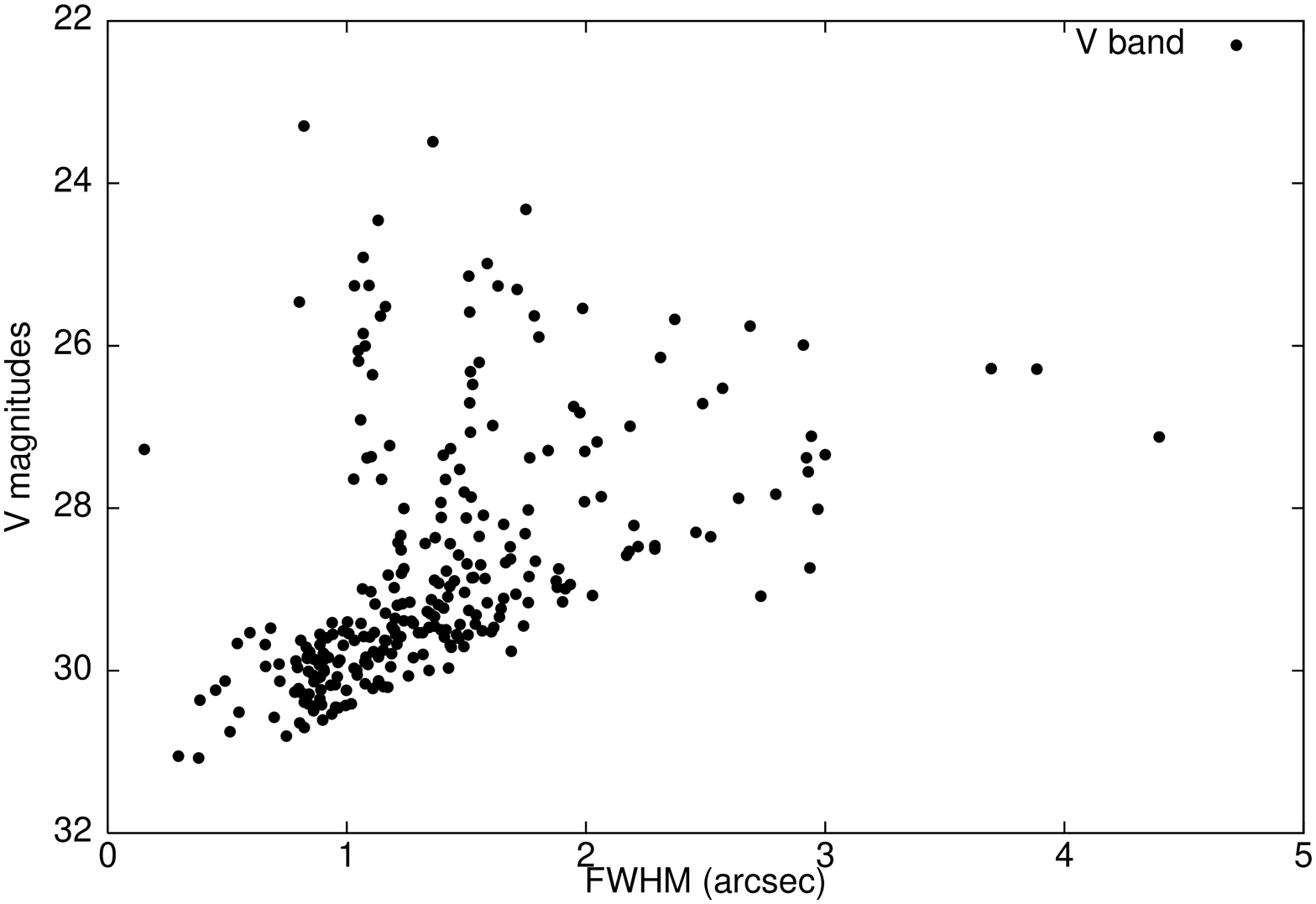}}
 \epsfxsize=5.2 cm \rotatebox{-90}{\epsffile{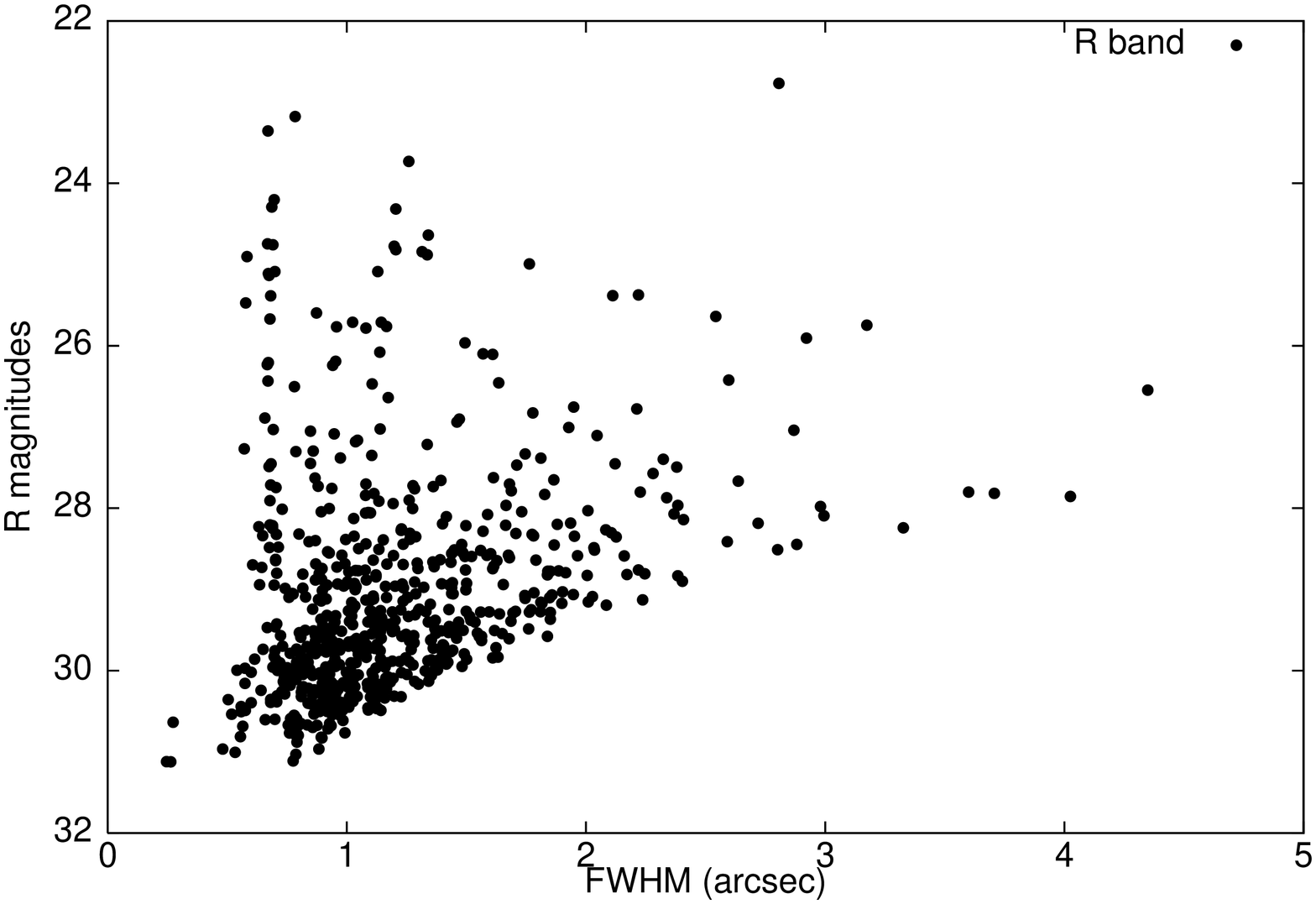}}
 \caption[]{All the objects detected in V (upper panel) and 
 R (lower panel) bands. A vertical stellar locus is
 present in both plots at the position of the seeing for each image. The
 magnitudes are not calibrated.}
 \label{allcat}
\end{figure}
We use the SExtractor\footnote{available at
http://terapix.iap.fr/soft/sextractor/index.html}
(Source-Extractor) package to build the catalogue for images V and
R. First we extract all the objects detected in both images with a
detection threshold of $2\sigma$ over the local sky. We show in
Fig.~\ref{allcat} all the detected objects in both bands,
representing uncalibrated magnitude vs. FWHM. In the two plots a
vertical stellar locus is clearly seen at the position of the
expected seeing for each image (FWHM$=1.1$ for V and FWHM$=0.75$
for R). We consider all objects to the right of these values as
being galaxies. In the V band, many objects lie on the lower left
side of the vertical stellar locus. We think the problem is due to
the low S/N value in the final V image, built with only 3 original
frames.

We select the galaxies present in both images and calibrate the
magnitudes. For the calibration we use data from the SuperCOSMOS
Sky Survey\footnote{http://www-wfau.roe.ac.uk/sss/} (SSS). We
obtain from the SSS the magnitudes in R and B$_J$ for two galaxies
in our field (see both marked in Fig.~\ref{Rgal}). The calibration
for our R filter is straightforward. For our V filter we use the
B$_J$ contained in the SSS. This means that our V filter is not
perfectly calibrated, but the offset does not induce any
difference in our results (since we are interested in the
shape/slope of the colour-magnitude diagram of our galaxies, the
offset induces only a vertical shift of all the objects in the
plot).

\begin{figure}[hbtp]
 \centering
 \epsfxsize=5.2 cm \rotatebox{-90}{\epsffile{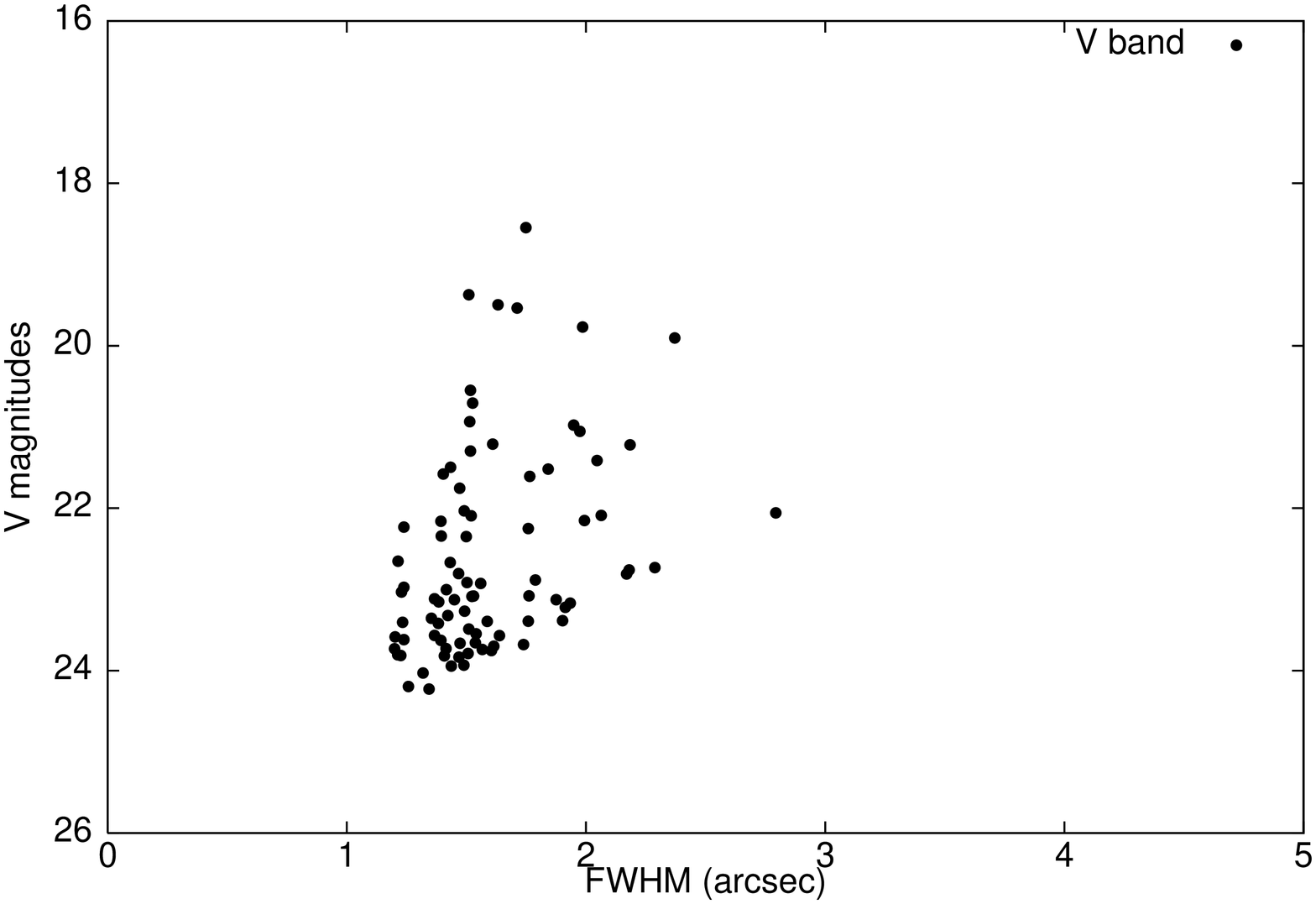}}
 \epsfxsize=5.2 cm \rotatebox{-90}{\epsffile{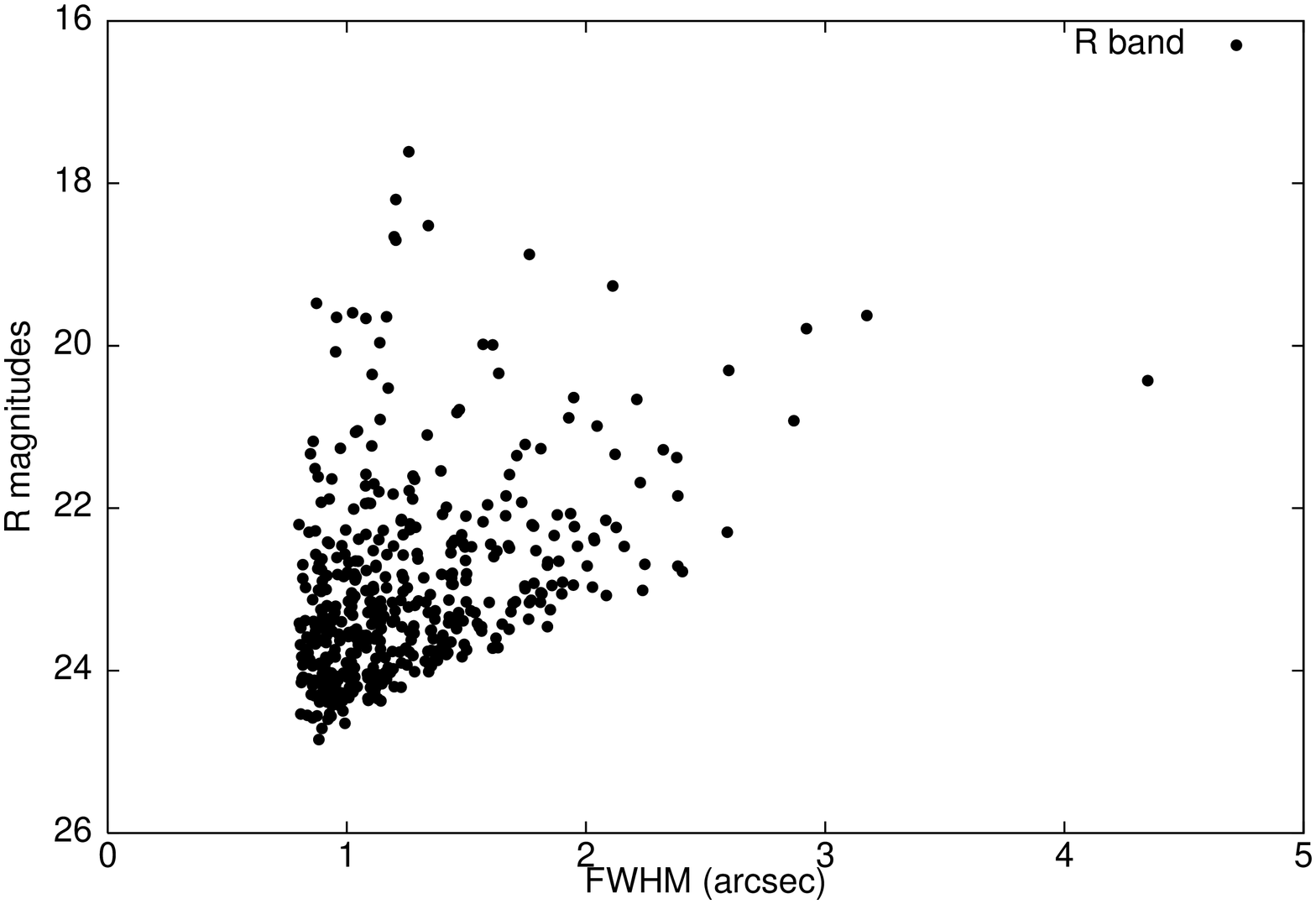}}
 \caption[]{Galaxies detected in V (upper panel) and R (lower panel)
 bands. The magnitudes are calibrated using the SSS archive.}
 \label{galcat}
\end{figure}
In Fig.~\ref{galcat} we show the selected galaxies in both V and R
images. Stars and deficient detections (SExtractor indicates this
with different flags) are rejected. The number of galaxies is 452
in the R filter and only 64 in the V filter. This represents a
70\% of the total number of objects detected in R and only a 23\%
of the objects detected in V.

The next step is to cross-check which galaxies detected in the V
image were also detected as galaxies in the R image. We find that
all the galaxies in V (64) are also present in the R catalogue.

\begin{figure}[hbtp]
 \centering
 \epsfxsize=5.2 cm \rotatebox{-90}{\epsffile{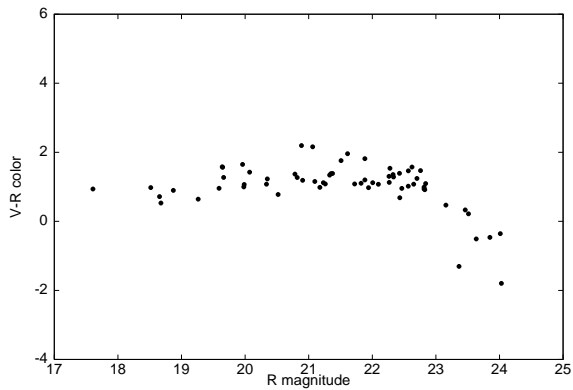}}
 \caption[]{Colour-magnitude diagram for the detected galaxies both in V and R
 filters. Although the final number of galaxies is low due to the low number of 
 detection in V band, a close relation can be infered at least up to a limit of 
 23$^{rd}$ magnitude.}
 \label{colormag}
\end{figure}

\begin{figure}[hbtp]
 \centering
 \includegraphics[bbllx=60,bblly=197,bburx=550,bbury=591,width=6.5cm,
                  angle=0,clip=true]{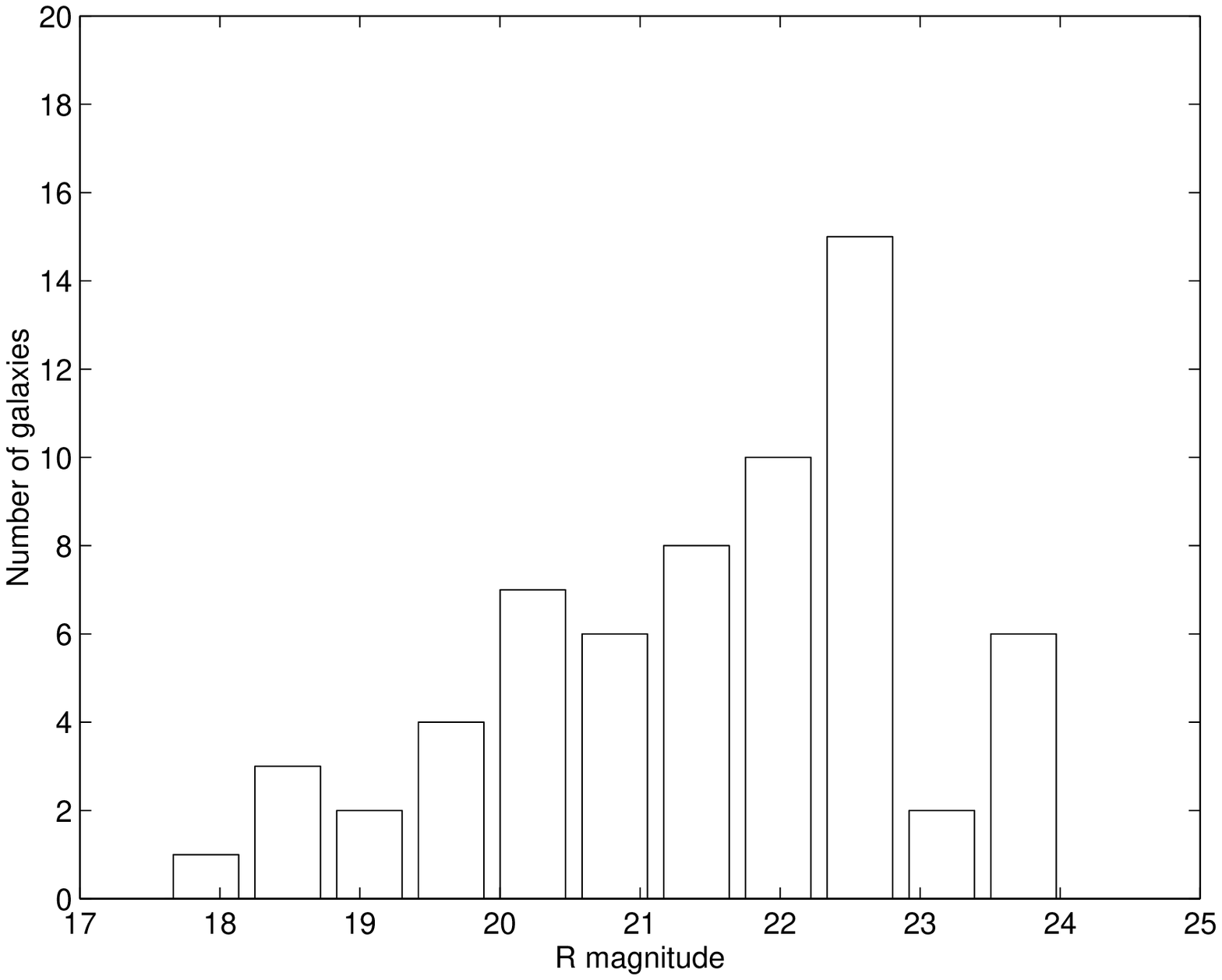}\\ 
 \includegraphics[bbllx=60,bblly=197,bburx=550,bbury=591,width=6.5cm,
                  angle=0,clip=true]{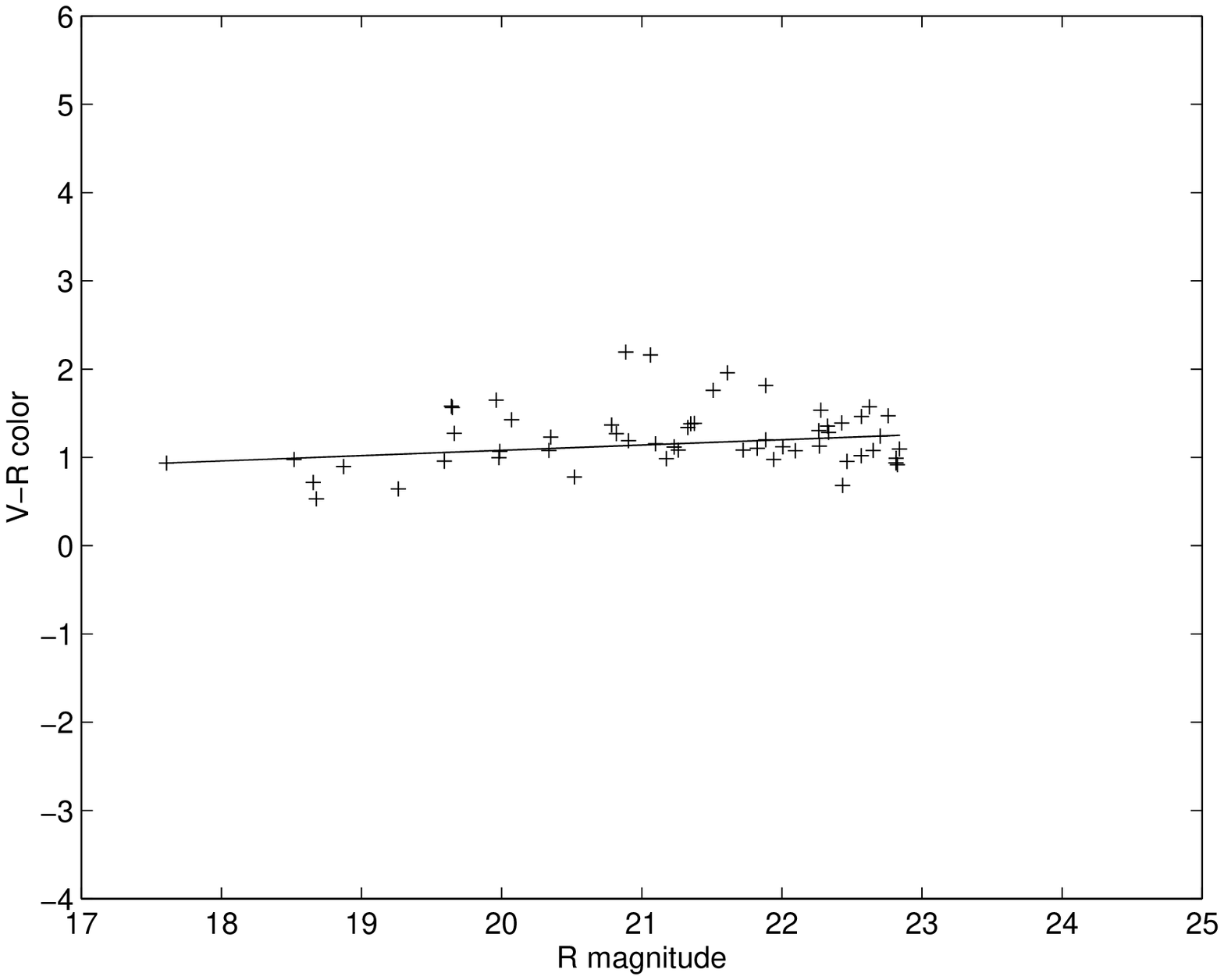}
 \caption[]{\emph{Upper panel}: In the distribution of galaxies for the R filter we see
 the completeness until the 23$^{rd}$ magnitude, where there is a drop in the number of
 galaxies detected. \emph{Lower panel}: The fit shows a red sequence of the detected
 early-type galaxies with a slope of 0.06.}
 \label{fit}
\end{figure}

The existence of a relation between colour and magnitude for
early-type galaxies is well known (Baum 1959; Sandage \&
Visvanathan 1978). In Fig.~\ref{colormag} we show the
colour-magnitude relation for the selected galaxies. Since the
presence of a red sequence of early-type galaxies is an almost
universal signature in clusters (Gladders et al. 1998, Gladders \&
Yee 2000 and references therein) and clusters at $z\approx0.5$
tend to concentrate elliptical galaxies in their central regions
(Dressler et al. 1997), we look for this sequence in our data. We
select only the galaxies below 23$^{rd}$ magnitude as this is our
completeness limit (see Fig.~\ref{fit} upper panel for
completeness), and we fit the remaining galaxies by a straight
line. Note that this fit is not sensitive to calibration problems,
these induce only a vertical shift in the line. We used a robust
statistical method based on minimizing the absolute deviation,
which is expected to be less sensitive to outliers compared to
standard linear regression (Press et al. 1992).

The result, presented in Fig.~\ref{fit} lower panel, shows a red
sequence with slope 0.06. According to the predicted slopes for
formation models of galaxy clusters as a function of redshift in
Gladders et al. (1998, see their Fig.~4), this slope is compatible
with a galaxy cluster at $z=0.5$. This value does not strongly 
depend on the cosmology.
This is particulary interesting because we would have
derived a most likely redshift of $\approx0.5$ from this
prediction, which is in good agreement with the actual redshift of
0.52.
\section{Comparison X-ray vs. Optical}

\begin{figure}[hbtp]
 \centering
 \epsfxsize=6.5 cm \rotatebox{-90}{\epsffile{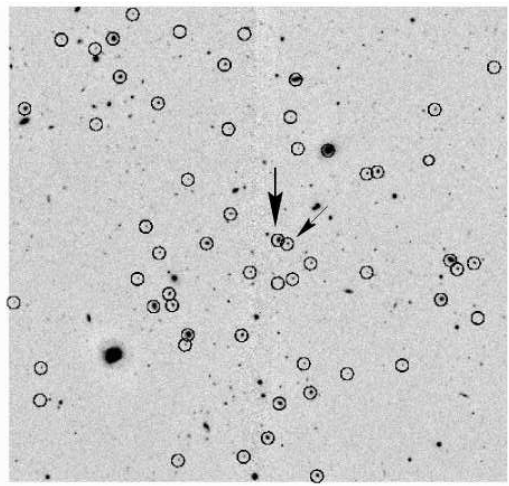}}
 \caption[]{Optical R band image of RBS380 (z=0.52). The total area is
 $5'\times5'$. We have inverted colours and marked the galaxies that were detected
 as cluster members using both R and V bands with a circle. The arrows indicate
 the two galaxies used for the calibration from the SSS (see text for details).
 The thicker arrow shows the AGN described in Sect.~\ref{s_xresults} and in
 Table~\ref{agntabla}. North is up and East is left.}
 \label{Rgal}
\end{figure}

In Fig.~\ref{Rgal} we show the selected galaxies through the colour-magnitude
relation, using the R image. We now want to compare the galaxy number density to
the distribution of the X-ray emission in the same area. For the number density
map, using a blank image of the same size as the optical image, we allocate 
pixels with value 1 in all the positions where we detected a
galaxy, and then we smooth it strongly (i.e. with a 200 pixels Gaussian). We
need such a large smoothing Gaussian because of the low number of galaxies
finally detected. In this way
we obtain the smooth distribution of the galaxies in the field.

\begin{figure}[hbtp]
 \centering
 \includegraphics[bbllx=105,bblly=85,bburx=487,bbury=449,width=6.5cm,
                  angle=0,clip=true]{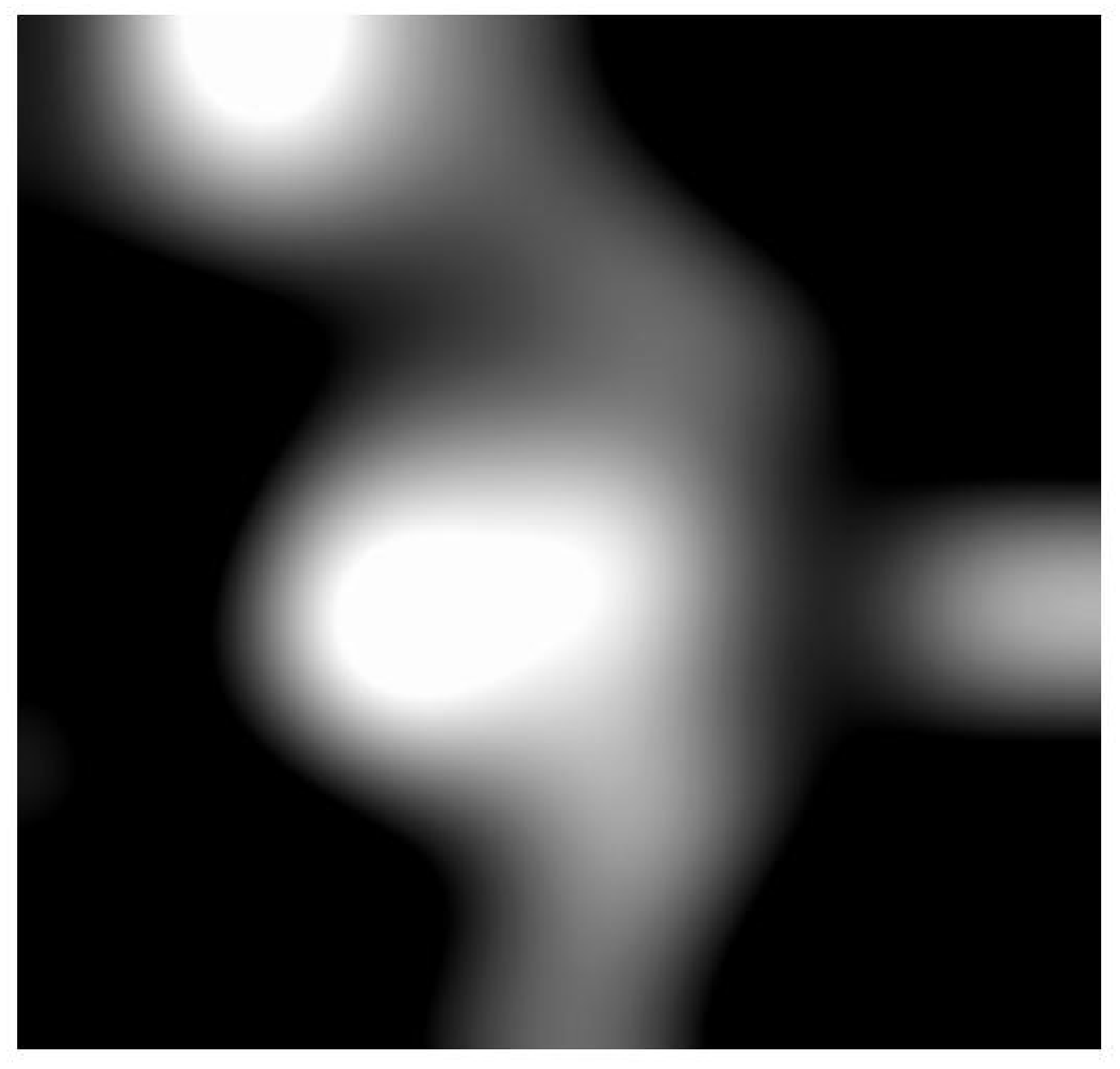}\\
 \includegraphics[bbllx=316,bblly=332,bburx=443,bbury=458,width=6.5cm,
                  angle=0,clip=true]{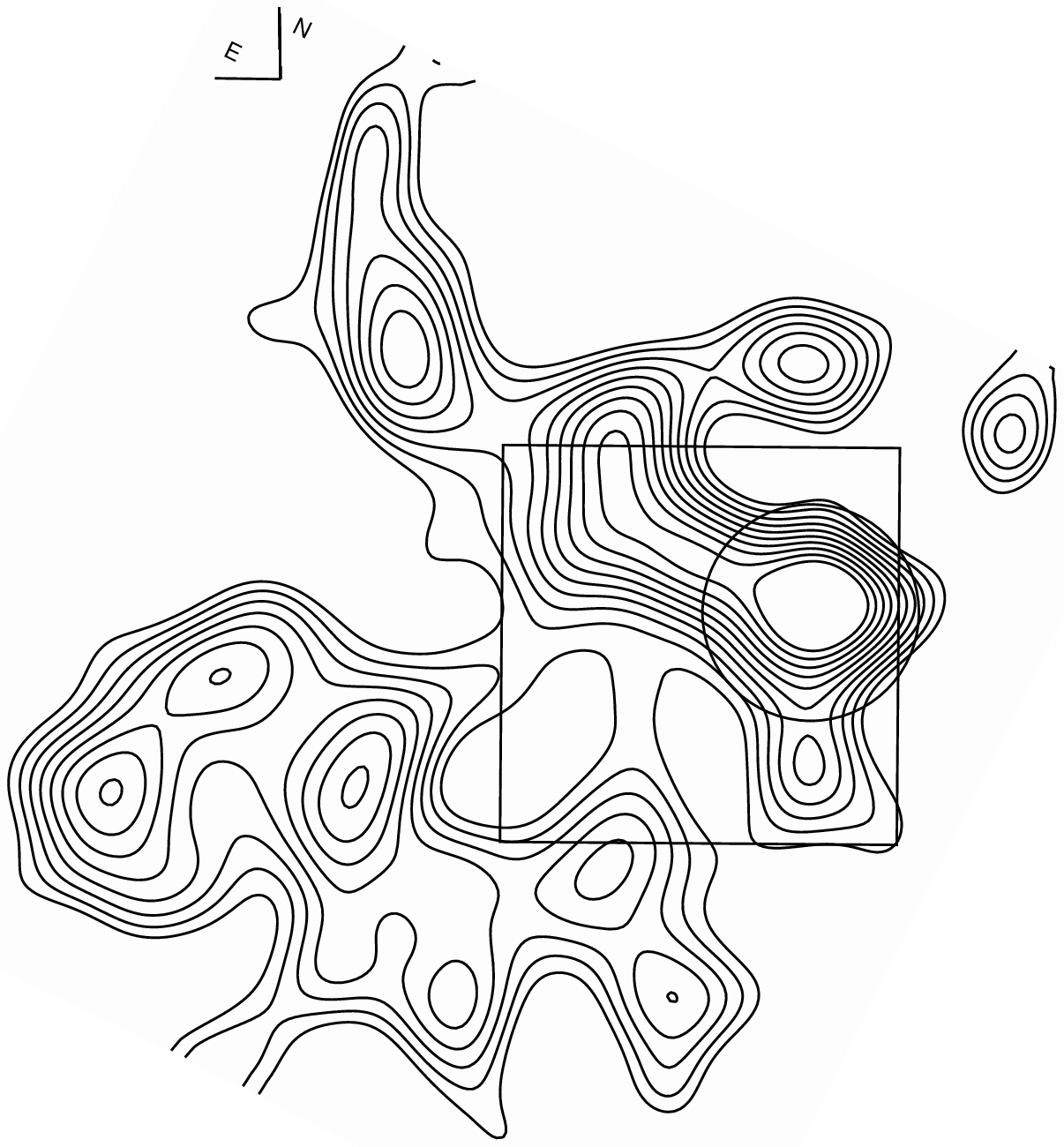}
 \caption[]{RBS380 galaxy number density in the R band (upper panel) and the
 X-rays contours for the same region (lower panel). The circle with radius $1'.5$ is
 the same as in Fig.~\ref{Xcumulo}. The total area in both panels is
 $5'\times5'$. North is up and East is left.}
 \label{comparison}
\end{figure}

From the X-ray image we extracted the contour lines from the
squared region shown in Fig.~\ref{Xcumulo} (which corresponds to
the observed region in the optical). In Fig.~\ref{comparison} we
plot the galaxy number density together with the X-ray contour
lines. The main maximum peak in the number density map is shifted
by $~2$ arcmin in SE direction with respect to the X-ray maximum.
Nevertheless, galaxies are present close to the asymmetric X-ray
features on both sides of the main peak (in N and NE direction).
These asymmetric features might indicate the existence of
surrounding material interacting with the cluster, e.g. infalling
galaxy groups.

The number of galaxies to the limiting magnitude is at least $2$ times
higher than expected for a such faint X-ray cluster (using the 
number of cluster members detected in an Abell radius of $R\leq1.5~h^{-1}$ 
within the centre of the cluster) but since the detection members 
efficiency is not complete due to the V band poor quality, this number 
could even be higher. This is another
confirmation that number of galaxies and X-ray luminosity are not well
correlated (see Table 1 for a comparison with other X-ray underluminous
clusters).

\section{Conclusions}

The X-ray source RBS380 was found in the RASS and identified
as a cluster of galaxies in the RBS. From the
RBS catalogue, the cluster was expected to be very massive 
due to its inferred high X-ray luminosity. Its redshift $z=0.52$
makes it the most distant galaxy cluster in that catalogue. Our
interest in this object was due to its predicted probability (up
to 60\%) of acting as a gravitational lens. In fact these
observations are part of a broader project that searches
systematically for gravitational arcs in different galaxy clusters
and combines this optical information with X-ray studies
of the same clusters in order to constrain cosmological models and
find possible correlations between X-ray and optical properties of
them.

With the new CHANDRA imaging we detect a strong X-ray point source
(an AGN) very close to the cluster centre, which could not be
resolved with ROSAT. After subtracting the emission of this AGN, 
the remaining diffuse emission is almost one order
of magnitude less luminous than expected: $L_X=1.6~10^{44}$ erg/s.
No previous investigation of the system has been carried out, so
our first aim was to make sure that it is really a cluster of
galaxies. The X-ray CHANDRA
observation shows a non-relaxed cluster of galaxies probably
interacting with surrounding material or/and another nearby
cluster.

From the NTT optical observations we are able to distinguish some
of the cluster members by means of the
colour-magnitude relation for early-type galaxies present in the
cluster, which is a well known signature for almost every cluster
of galaxies. The obtained slope for this red sequence is 0.06.
Using existing predicted slopes for different formation models as
a function of redshift, the most likely redshift for this slope is
$z\approx0.5$, in good agreement with the measured redshift of
$0.52$.

We could not detect any gravitational arc in this cluster. This is
not surprising as with the low X-ray luminosity the
probability for arcs is strongly reduced.

The example of this cluster shows that high-resolution X-ray
imaging is crucial for cosmological research. This type of distant
galaxy clusters is often used for various types of cosmological
applications. Due to source confusion some clusters can have wrong
luminosity measurements and hence influence the results. This effect
might e.g. artificially flatten the luminosity function for
distant clusters.

\begin{table}[tb]
 \centering
 \begin{tabular}{lllll}
  \hline\noalign{\smallskip}
   Name & Redshift & Luminosity [erg/s] & band\\
  \noalign{\smallskip}\hline\noalign{\smallskip}
   Cl0500$-$24   & 0.32 & 5.6~10$^{44}$  & bolometric \\
   Cl0939$+$4713 & 0.41 & 7.9~10$^{44}$  & bolometric \\
   RBS380        & 0.52 & 2  ~10$^{44}$  & bolometric \\
  \noalign{\smallskip}\hline
 \end{tabular}
 \caption{We compare the X-ray luminosity of RBS380 with two more clusters of
  galaxies which are optically rich, but have relatively low X-ray
  luminosity. For comparison, we give the bolometric luminosity for RBS380
  too. }
\label{cuadro}
\end{table}
\begin{acknowledgements}

This work has been partly supported by a predoctoral Marie Curie
Fellowship to RGM at the Liverpool John Moores Astrophysics
Research Institute (Marie Curie Training Site HPMT-CT-2000-00136), by a
German Science Foundation grant (WA~1047/6-1) and
by the Austrian Science Foundation (FWF P15868).
We thank Olivier Hainaut (ESO NTT-Team) for detailed explanations
on hyper-flatfielding. We are grateful to Joachim Wambsganss and
Axel Schwope for their work in the arc search programme, by which
programme the data sets were provided. We also thank Eelco van Kampen
for a number of simulations for better understanding our CM diagrams and 
Robert Schmidt for comments to a first version of this paper.
RGM specially thanks Elisabetta De Filippis and Africa Castillo-Morales 
for time-sharing and useful discussions during this work and the Institute
f\"{u}r Astrophysik in Innsbruck for its hospitality.

\end{acknowledgements}

{}

\end{document}